\documentstyle[12pt]{article}
\newcommand{\journal}[4]{{\em #1~}#2\,(19#3)\,#4;}

\newcommand{\ijmp}{\journal {Int. J. Mod. Phys.}}

\newcommand{\pr}{\journal {Phys. Rev.}}

\newcommand{\prl}{\journal {Phys. Rev. Lett.}}
\newcommand{\jmp}{\journal {J. Math. Phys.}}

\newcommand{\cmp}{\journal {Comm. Math. Phys.}}
\newcommand{\cqg}{\journal {Class. Quantum Grav.}}

\newcommand{\np}{\journal {Nucl. Phys.}}
\newcommand{\pl}{\journal {Phys. Lett.}}

\newcommand{\nc}{\journal {Nuovo Cim.}}

\makeatletter
\@addtoreset{equation}{section}
\makeatother

\def\LP{\displaystyle{\Biggl(}}
\def\RP{\displaystyle{\Biggr)}}

\newcommand{\G}{\Gamma}
\newcommand{\D}{\Delta}
\newcommand{\K}{\widetilde{K}}
\renewcommand{\L}{\widetilde{L}}
\newcommand{\M}{\widetilde{M}}
\newcommand{\N}{\widetilde{N}}
\newcommand{\U}{\widetilde{U}}
\newcommand{\V}{\widetilde{V}}
\newcommand{\wT}{\widetilde{T}}

\newcommand{\vr}{\varrho}

\renewcommand{\o}{\omega}

\newcommand{\s}{\sigma} \renewcommand{\S}{\Sigma}

\renewcommand{\AA}{{\cal A}}
\newcommand{\BB}{{\cal B}}

\newcommand{\DD}{{\cal D}}

\newcommand{\FF}{{\cal F}}
\newcommand{\GG}{{\cal G}}
\newcommand{\HH}{{\cal H}}

\newcommand{\JJ}{{\cal J}}

\newcommand{\NN}{{\cal N}}

\newcommand{\RR}{{\cal R}}
\newcommand{\SS}{{\cal S}}
\newcommand{\TT}{{\cal T}}
\newcommand{\UU}{{\cal U}}

\newcommand{\complex}{{\kern .1em {\raise .47ex
\hbox {$\scriptscriptstyle |$}}
    \kern -.4em {\rm C}}}
\newcommand{\real}{{{\rm I} \kern -.19em {\rm R}}}
\newcommand{\rational}{{\kern .1em {\raise .47ex
\hbox{$\scripscriptstyle |$}}
    \kern -.35em {\rm Q}}}
\renewcommand{\natural}{{\vrule height 1.6ex width
.05em depth 0ex \kern -.35em {\rm N}}}

\newcommand{\cb}{{\bar c}}
\newcommand{\ob}{{\bar \omega}}
\newcommand{\vfb}{{\bar \varphi}}

\newcommand{\pad}[2]{{\frac{\partial #1}{\partial #2}}}
\newcommand{\fud}[2]  {{\displaystyle{\frac{\delta #1}{\delta #2}}}}

\newcommand{\sla}{\raise.15ex\hbox{$/$}\kern -.57em}

\newcommand{\twiddle}{\lower.9ex\rlap{$\kern -.1em\scriptstyle\sim$}}

\newcommand{\vf}{{\varphi}}

\newcommand{\equ}[1]{(\ref{#1})}

\newcommand{\eq}{\begin{equation}}
\newcommand{\eqn}[1]{\label{#1}\end{equation}}
\newcommand{\eea}{\end{eqnarray}}
\newcommand{\eqa}{\begin{eqnarray}}
\newcommand{\eqan}[1]{\label{#1}\end{eqnarray}}
\newcommand{\ba}{\begin{array}}
\newcommand{\ea}{\end{array}}
\newcommand{\eqac}{\begin{equation}\begin{array}{rcl}}
\newcommand{\eqacn}[1]{\end{array}\label{#1}\end{equation}}

\renewcommand{\pad}[2]{{\displaystyle{\frac{\partial #1}{\partial #2}}}}

\newcommand{\intx}{\int d^4 \! x \, }

\begin{document}
\def\ftoday{{\sl  \number\day \space\ifcase\month
\or Janvier\or F\'evrier\or Mars\or avril\or Mai
\or Juin\or Juillet\or Ao\^ut\or Septembre\or Octobre
\or Novembre \or D\'ecembre\fi
\space  \number\year}}
%
%
\titlepage

\begin{center}

{\huge Landau Gauge within the Gribov Horizon}

\vspace{1cm}

{\Large Nicola Maggiore\footnote{Supported by Universit\`a degli Studi
di Genova and Della Riccia Foundation. \\Present address~: D\'epartement de
Physique Th\'eorique, Universit\'e de Gen\`eve, CH -- 1211 Gen\`eve 4,
Switzerland. e-mail~: maggiore@sc2a.unige.ch} and Martin
Schaden\footnote{Supported by Deutsche
Forschungsgemeinschaft. e-mail~: schaden@mafalda.physics.nyu.edu}}

\vspace{.5cm}

{\it
Physics Department, New York University\\4 Washington Place,
 New York, N.Y. 10003}

\end{center}

\vspace{1.7cm}

\begin{center}
\bf ABSTRACT
\end{center}
{\it
We consider a model which effectively restricts the functional integral
of Yang--Mills theories to the fundamental modular region. Using
algebraic arguments, we prove that this theory has the same divergences
as ordinary Yang Mills theory in the Landau gauge and that it is unitary.
The restriction of the functional integral is interpreted as a
kind of spontaneous breakdown of the $BRS$ symmetry.}

\vfill

NYU--TH--93/10/05\hfill October 1993
\newpage

\section{Introduction}

A perturbative expansion of gauge field theories requires gauge fixing. In
1978 Gribov~\cite{gribov} pointed out that
covariant gauges (and most others that can be cast in the form of a local
effective action) do not uniquely specify a single configuration on a gauge
orbit in nonabelian theories. This ambiguity can be
disregarded when discussing high-energy processes and does not affect the
well known perturbative results of asymptotic freedom, or equivalently,
asymptotic scaling. At low energies, the presence of additional gauge copies
can however no longer be overlooked~\cite{gribov}, and this is at least one
reason why a perturbative analysis is bound to fail in this case.

Gribov~\cite{gribov} suggested to restrict
the functional integration to  the space of configurations $A$
which are transverse
\eq
\partial A = 0\ ,
\eqn{111}
and such that
\eq
\partial D(A) \leq 0\ ,
\eqn{112}
where $D(A)$ is the covariant derivative. The boundary of the region so
defined is called the Gribov horizon and lies inside a certain
ellipsoid~\cite{Dantonio}.

The restriction to the region defined by~\equ{111} and~\equ{112} would
already imply that the
gluon propagator differs from the usual one at low momenta~\cite{gribov}
\eq
D^{ab}_{\mu\nu}=\delta^{ab}(\delta_{\mu\nu}-\frac{k_\mu k_\nu}{k^2})
\frac{k^2}{k^4 +g^2N\gamma^2}\ ,
\eqn{18}
and depends on a dimensionful parameter $\gamma$.

The Fundamental Modular Region (FMR), where the Hilbert norm of the connection
\eq
F_A(U) \equiv ||A^U||^2=||U^\dagger AU+U^\dagger dU||^2\ ,
\eqn{113}
attains its absolute minimum
with respect to gauge transformations $U$~\cite{fmr,zw82}
\eq
\mbox{FMR}\equiv\{A:F_A(1)\leq F_A(U)\}\ ,
\eqn{114}
is a proper subset of the region defined by Gribov, since~\equ{111}
and~\equ{112} characterize any relative minimum of the Hilbert norm~\equ{113},
and such relative minima have been shown to exist~\cite{minima}. The FMR is
therefore a refinement of Gribov's region and in particular of the definition
of the Landau gauge~\equ{111} and has for this reason been called ``minimal''
Landau gauge~\cite{zw82}. Zwanziger proposed a local action which concretely
implements the restriction to the FMR when a certain nonperturbative
``horizon condition''~\cite{zw89,zw92} is satisfied.
He studied this mechanism in the
continuum~\cite{zw89} as well as in the critical limit of lattice gauge
theory~\cite{zw92,zwpreprint}. Quite
remarkably this version of the $SU(N)$ Yang-Mills
theory naturally gives a  gluon propagator of the Gribov type~\equ{18} where
 the parameter $\gamma$ is selfconsistently determined by the
``horizon condition''~\cite{zw89,zw92}.

The required locality of the classical action immediatly
raises the question of renormalizability: whether new divergences or
anomalies not present in the original YM-theory have been introduced,
which would imply that the proposed gauge-fixing is  not
renormalizable.
In Ref.~\cite{zw93} an analysis based on the BRS symmetry of the model
indicated that radiative corrections develop at most four divergences.
Arguments were put forward that only two of these should occur in a
perturbative expansion.

The Landau gauges, the restriction to the FMR being the ``minimal'' one,
have well known nonrenormalization properties~\cite{book}. In this paper
we algebraically recover these properties for the new gauge model, namely
that  only two independent renormalization constants are needed, and thus
 confirm also from the renormalization point of view that the model
belongs to the class of Landau gauge theories. We also prove that the
restriction to the FMR does not spoil the unitarity of the theory.

The paper is organized as follows.
In Sect.~2 we present the symmetries of the model, which
include a Ward identity typical for the Landau gauge~\cite{landau}. The
full algebraic structure allows us to give in Sect.~3 a formal proof
that indeed only two divergences are present. At the same time we
demonstrate that the whole algebra is free of anomalies. Sect.~4 is
devoted to an interpretation of the model at the nonvanishing physical
value for the external sources. We show that this is equivalent to a kind of
spontaneous breakdown of the $BRS$ symmetry. The results are then summarized
in Sect.~5.

\section{The model and its algebraic structure}

To constrain the functional integral to the FMR, additional fields
$(\omega^a_i, \varphi^a_i, \bar\omega^{ai}, \bar\varphi^{ai})$ and
external sources $(U^{ai}_\mu,V^{a}_{\mu i}, M^{ai}_\mu, N^{a}_{\mu i})$
were introduced~\cite{zw93} into the original Yang--Mills theory. These new
fields transform under a global $U(f)$ symmetry on the composite
index~\mbox{$i=(\mu, a)$},  with $f=4(N^2-1)$.

The model in Euclidean space--time is described by the action~\cite{zw93}
\eq
S = S_{LYM} + s\intx  \LP (\partial_\mu\bar\omega^{ai}) (D_\mu\varphi_i)^a +
U^{ai}_\mu(D_\mu\varphi_i)^a + V^{a}_{\mu i}(D_\mu\ob^i)^a +U^{ai}_\mu
V^{a}_{\mu i}\RP\ ,
\eqn{11}
where $S_{LYM}$ is the ordinary Yang Mills action in the Landau gauge
\eq
S_{LYM}=\frac{1}{4g^2}\intx F^{a}_{\mu\nu}F^a_{\mu\nu} - s
\intx{(\partial_\mu\bar
c}^a)A_\mu^a\ ;
\eqn{12}
\eq
F^a_{\mu\nu}=\partial_\mu A^a_\nu - \partial_\nu A^a_\mu
+f^{abc}A^b_\mu A^c_\nu\ ,
\eqn{13}
and the covariant derivative is defined as
\eq
(D_\mu X)^a = \partial_\mu X^a + f^{abc}A^b_\mu X^c\ ,
\eqn{14}
with $f^{abc}$ being the structure constants of the gauge group.
Finally, the operator $s$ is the ordinary $BRS$ transformation
extended
to the additional fields
\eq\ba{rcl}
s A^{a}_{\mu}{\ } =& - {(D_{\mu}c)}^{a}&   \\
s c^{a}{\ }{\ } =& {1 \over 2}f^{abc}c^b c^c&  \\
s {\bar c}^{a}=& b^{a}\hfil & s b^{a}=0  \\
s{\bar \omega}^{ai}=& {\bar \varphi}^{ai}\hfil & s{\bar\varphi}^{ai}=0\\
s\varphi^a_i=& \omega^a_i\hfil & s\omega^a_i=0\ ,
\ea
\eqn{15}
and to the sources by
\eq\ba{lcl}
s U^{ai}_\mu=M^{ai}_\mu && s M^{ai}_\mu=0\\
s V^a_{\mu i}=N^a_{\mu i} && s N^a_{\mu i}=0\ .
\ea
\eqn{16}
One can easily verify the nilpotency of the
$BRS$ operator~\equ{15},~\equ{16}
\eq
s^2=0\ .
\eqn{23}
In Ref.~\cite{zw93} it is argued that the physical value for the sources is
\eq
\begin{array}{lcl}
N^{ab}_{\mu\nu}=U^{ab}_{\mu\nu}&=&0\\
M^{ab}_{\mu\nu}=-V^{ab}_{\mu\nu}&=&\gamma\delta_{\mu\nu}\delta^{ab}\ ,
\end{array}
\eqn{17}
where $\gamma$ is a parameter of dimension $[{\rm mass}]^2$, whose value
is determined by a selfconsistency condition that will be discussed in
Sect. 4.
The $BRS$ symmetry~\equ{15}--~\equ{16} is the simplest which is
cohomologically equivalent to the ordinary one, because the additional
fields transform as doublets~\cite{dixon}.
With the $BRS$ transformation~\equ{15}--~\equ{16}, the
action corresponding to~\equ{11} explicitely is~\footnote[1]{Our conventions
differ from those of reference~\cite{zw93} and the $\vf$--$\vfb$
propagator of~\equ{24} is positive.}
\eq\ba{rl}
S =\intx\LP & \frac{1}{4g^2}F^{a}_{\mu\nu}F^a_{\mu\nu}
-(\partial_\mu b^a) A^a_\mu -(\partial_\mu{\bar c}^a) (D_\mu c)^a\\
&
+(\partial_\mu\bar\varphi^{ai}) (D_\mu \varphi_i)^a
-(\partial_\mu\bar \omega^{ai}) (D_\mu \omega_i)^a
+f^{abc}(\partial_\mu \ob^{ai})(D_\mu c)^b \varphi_i^c\\
&
+ M^{ai}_\mu (D_\mu \varphi_i)^a
- U^{ai}_\mu (D_\mu \omega_i)^a
+ f^{abc}U^{ai}_\mu (D_\mu c)^b \varphi_i^c\\
&+ N^a_{\mu i} (D_\mu \bar\omega^i)^a
+ V^a_{\mu i} (D_\mu \bar\varphi^i)^a
-f^{abc}V^a_{\mu i}(D_\mu c)^b\ob^{ci}
+ M^{ai}_\mu V^a_{\mu i}
- U^{ai}_\mu N^a_{\mu i}\RP
\ ,
\ea\eqn{24}
where the dimension and ghost charge assignments of the fields are
summarized in Table~1.

As customary, we couple external sources to the nonlinear $BRS$
variations in~\equ{15} of the quantum fields. The full classical action
\eq
\Sigma=S+S_{ext}\ ,
\eqn{31}
with
\eq
S_{ext}=\intx\LP\!\! K^a_\mu (sA^a_\mu) + L^a (sc^a) \RP\ ,
\eqn{32}
then satisfies the Slavnov identity
\eq
{\cal S}(\Sigma) = 0                           \ ,
\eqn{33}
where
\eq\ba{rl}
{\cal S}(\S)  =
\intx \LP &\!\!
       \fud{\S}{K^a_\mu} \fud{\S}{A^{a}_{\mu}}
   +   \fud{\S}{L^{a}} \fud{\S}{c^{a}}
   +    b^a\fud{\S}{{\bar c}^{a}}
   +    \omega^a_i \fud{\S}{\vf^a_i}
   +    \vfb^{ai}\fud{\S}{\ob^{ai}}
\\&
   +    M^{ai}_\mu \fud{\S}{U^{ai}_\mu}
   +    N^a_{\mu i} \fud{\S}{V^a_{\mu i}}\RP\ ,
\ea
\eqn{34}
and the corresponding linearized operator
\eq\ba{rl}
\BB_\S =   \intx \LP &\!\!
       \fud{\S}{K^a_\mu} \fud{\ }{A^{a}_{\mu}}
+       \fud{\S}{A^a_\mu} \fud{\ }{K^{a}_{\mu}}
   +   \fud{\S}{L^{a}} \fud{\ }{c^{a}}
   +   \fud{\S}{c^{a}} \fud{\ }{L^{a}}
   +    b^a\fud{\ }{{\bar c}^{a}}
\\&
   +    \omega^a_i \fud{\ }{\vf^a_i}
   +    \vfb^{ai}\fud{\ }{\ob^{ai}}
   +   M^{ai}_\mu \fud{\ }{U^{ai}_\mu}
   +   N^a_{\mu i} \fud{\ }{V^a_{\mu i}}\RP\ ,
\ea\eqn{35}
is nilpotent
\eq
\BB_\S \BB_\S=0\ .
\eqn{36}
For theories in the Landau gauge, the integrated ghost equation of
motion gives a Ward identity~\cite{landau}, which in our case is
\eq
{\cal G}^{a} \S=\Delta^a\ ,
\eqn{26}
where
\eq
{\cal G}^{a} = \intx {\ }\left({\ }
    { \delta {\ } \over \delta c^{a} }
   + f^{abc}\LP {\bar c}^{b}{ \delta {\ } \over \delta b^{c} }
+ \vf^b_i\fud{\ }{\omega_i^c}
+ \ob^{bi}\fud{\ }{\vfb^{ci}}
+ V^b_{\mu i}{ \delta {\ } \over \delta N^c_{\mu i}}
+ U^{bi}_\mu{ \delta {\ } \over \delta M^{ci}_\mu}
{\ }\right)\RP\ ,
\eqn{27}
and
\eq
\Delta^a=\intx f^{abc}\LP K^b_\mu A^c_\mu- L^bc^c\RP\
\eqn{38}
is a linear breaking in the quantum fields, thus only present at the
classical level.

The anticommutator
 between the ghost equation~\equ{26}
and the Slavnov identity~\equ{33} is known to give the Ward
identity of rigid gauge invariance~\cite{landau} (see~(2.33))
\eq
{\cal H}^a_{rig}\S=0\ ,
\eqn{39}
where
\eq
{\cal H}^a_{rig}=\sum_{(all{\ }fields{\ }\Phi)}\intx\
f^{abc}\Phi^b\frac{\delta}{\delta\Phi^c}\ .
\eqn{310}

We can also write the Ward identity
\eq
{\cal F}^i\Sigma=0\ ,
\eqn{311}
with
\eq
{\cal F}^i = \intx \LP
c^a\fud{\ }{\omega^a_i}
- \bar\omega^{ai}\fud{\ }{\bar{c}^a}
-U^{ai}_\mu\fud{\ }{K^a_\mu}
\RP\ .
\eqn{312}

Commuting ${\cal F}^i$ with the Slavnov operator~\equ{34},
another nonlinear symmetry emerges (see~(2.33))
\eq
{\cal J}^i(\Sigma)=0\ ,
\eqn{314}
with
\eq
{\cal J}^i(\S)=\intx\LP
     c^{a} \fud{\S}{{\vf^a_i}}
   -  \fud{\S}{L^{a}} \fud{\S}{\omega^a_i}
+\vfb^{ai}\fud{\S}{\cb^a}
   +    M^{ai}_\mu \fud{\S}{K^a_\mu}    \RP\ .
\eqn{315}

In addition we have the symmetry
\eq
{\cal R}_i^j\Sigma=0\ ,
\eqn{316}
where
\eq
{\cal R}_i^j=\intx \LP
\vf^a_i \fud{\ }{\omega^a_j}
+ V^a_{\mu i} \fud{\ }{N^a_{\mu j}}
-\bar\omega^{aj}\fud{\ }{\bar\vf^{ai}}
- U^{aj}_\mu \fud{\ }{M^{ai}_\mu}
\RP\ .
\eqn{317}
Anticommuting the symmetry~\equ{316} with the Slavnov
identity~\equ{33}
we get the Ward identity of the $U(f)$ symmetry (see~(2.33))
\eq
{\cal U}^j_i\Sigma=0\ ,
\eqn{319}
where
\eq
\ba{rl}
{\cal U}^j_i=\intx\LP &
\vf_i^a\fud{\ }{\vf^a_j}
+\omega_i^a\fud{\ }{\omega^a_j}
+V_{\mu i}^a\fud{\ }{V^a_{\mu j}}
+N_{\mu i}^a\fud{\ }{N^a_{\mu j}}
\\&
-\bar\vf^{aj}\fud{\ }{\bar\vf^{ai}}
-\bar\omega^{aj}\fud{\ }{\bar\omega^{ai}}
-U^{aj}_\mu\fud{\ }{U_\mu^{ai}}
-M^{aj}_\mu\fud{\ }{M_\mu^{ai}}
\RP\ .
\ea\eqn{320}
By means of the diagonal operator $Q_f={\cal U}^i_i$ the $i$--valued
fields are assigned an additional quantum number.

In Table 1 we summarize the quantum numbers of the fields
and sources.

\begin{center}
\begin{tabular}{|l|r|r|r|r|r|r|r|r||r|r|r|r|r|r|}\hline
&$A$ & $c$ & $\bar c$ & $b$ & $\omega$ & $\bar\omega$ &
$\varphi$ & $\bar\varphi$&$K$&$L$&$M$&$N$&$U$&$V$\\ \hline
dim&1&0&2&2&1&1&1&1&3&4&2&2&2&2\\ \hline
$\Phi\Pi$&0&1&$-1$&0&1&$-1$&0&0&$-1$&$-2$&0&1&$-1$&0\\ \hline
$Q_f$&0&0&0&0&1&$-1$&1&$-1$&0&0&$-1$&1&$-1$&1\\ \hline
\end{tabular}

\vspace{.2cm}{\footnotesize
{\bf Table 1.} Dimensions, Faddeev--Popov charges and $Q_f$ numbers of
the fields.}
\end{center}

Apart from the familiar gauge condition and antighost equation
\eq
\fud{\S}{b^a} - \partial_\mu A^a_\mu\nonumber = 0
\eqn{321}
\eq
\partial^\mu\fud{\S}{K^{a\mu}} + \fud{\S}{\bar{c}^a}= 0\ ,
\eqn{antighosteq}
this model is also characterized by the set of local equations
\eq
{\cal T}_{(\Phi)}\Sigma=\Delta_{(\Phi)}\ ,
\eqn{322}
where
\eqa
{\cal T}_{(\omega)}^{ai} &=& \fud{\ }{\omega^a_i}
+ \partial_\mu\fud{\ }{N^a_{\mu i}} + f^{abc}\ob^{bi}\fud{\ }{b^c}
\nonumber\\
{\cal T}_{(\vf)}^{ai} &=& \fud{\ }{\vf^a_i}
+ \partial_\mu\fud{\ }{V^a_{\mu i}}
+f^{abc}\LP \vfb^{bi}\fud{\ }{b^c}
+\ob^{bi}\fud{\ }{\cb^c}
+U^{bi}_\mu\fud{\ }{K^c_\mu}\RP
\nonumber\\
{\cal T}_{(\vfb)\,i}^a &=& \fud{\ }{\vfb^{ai}}+
\partial_\mu\fud{\ }{M^{ai}_\mu}
\nonumber\\
{\cal T}_{(\ob)\,i}^a &=& \fud{\ }{\ob^{ai}}+
\partial_\mu\fud{\ }{U^{ai}_\mu}
+f^{abc}V^b_{\mu i}\fud{\ }{K^c_\mu}
\nonumber\\
&&\\
\Delta_{(\omega)}^{ai} &=& f^{abc} U^{bi}_\mu A^c_\mu
\nonumber\\
\Delta_{(\vf)}^{ai} &=& f^{abc}M^{bi}_\mu A^c_\mu
\nonumber\\
\Delta_{(\vfb)\,i}^a &=& f^{abc}V^b_{\mu i}A^c_\mu
\nonumber\\
\Delta_{(\ob)\,i}^a &=&-f^{abc} N^b_{\mu i}A^c_\mu \nonumber \ .
\eea

In the proof of the renormalization of a model, the nonlinear algebra formed
by the symmetry operators plays an important role, because it yields
consistency conditions on the counterterm and on the possible anomalies
of the theory.

The only nontrivial algebraic relations are
\eq\ba{l}
 \BB_\Psi {\cal S}(\Psi) =0
\nonumber\\
 \fud{\ }{b^a}{\cal S}(\Psi)-
\BB_\Psi(\fud{\Psi}{b^a}-\partial_\mu A^a_\mu)=
\fud{\Psi}{\cb^a}+\partial_\mu\fud{\Psi}{K^a_\mu}
\nonumber\\
{\cal G}^a{\cal S}(\Psi)+
 \BB_\Psi({\cal G}^a\Psi-\Delta^a)={\cal H}^a_{rig}\Psi
\nonumber\\
{\cal F}^i{\cal S}(\Psi)-\BB_\Psi {\cal F}^i\Psi =
{\cal J}^i(\Psi)
\nonumber\\
{\cal T}_{(\vfb)\,i}^a{\cal S}(\Psi)-\BB_\Psi (
{\cal T}_{(\vfb)\,i}^a\Psi-\Delta_{(\vfb)\,i}^a)=
{\cal T}_{(\ob)\,i}^a\Psi-\Delta_{(\ob)\,i}^a
\nonumber\\
{\cal T}_{(\omega)}^{ai}{\cal S}(\Psi)+
\BB_\Psi({\cal T}_{(\omega)}^{ai}\Psi-\Delta_{(\omega)}^{ai})
={\cal T}_{(\vf)}^{ai}\Psi-\Delta_{(\vf)}^{ai}
\nonumber\\
{\cal G}^a {\cal F}^i\Psi
-{\cal F}^i
({\cal G}^a\Psi-\Delta^a)=
\intx({\cal T}_{(\omega)}^{ai}\Psi-\Delta_{(\omega)}^{ai})
\\
(\fud{\ }{\cb^a}+\partial_\mu\fud{\ }{K^a_\mu})
({\cal G}^b\Psi-\Delta^b)+
{\cal G}^b(\fud{\Psi}{\cb^a}+\partial_\mu\fud{\Psi}{K^a_\mu})
=-f^{abc}(\fud{\Psi}{b^c}-\partial_\mu A^c_\mu)
\nonumber\\
{\cal T}_{(\vf)}^{ai}({\cal G}^b\Psi-\Delta^b)-
{\cal G}^b ({\cal T}_{(\vf)}^{ai}\Psi-\Delta_{(\vf)}^{ai})=
-f^{abc}({\cal T}_{(\omega)}^{ci}\Psi-\Delta_{(\omega)}^{ci})
\nonumber\\
{\cal T}_{(\ob)\,i}^a({\cal G}^b\Psi-\Delta^b)+
{\cal G}^b ({\cal T}_{(\ob)\,i}^a\Psi-\Delta_{(\ob)\,i}^a)=
-f^{abc}({\cal T}_{(\vfb)\,i}^c\Psi-\Delta_{(\vfb)\,i}^c)
\nonumber\\
{\cal G}^a{\cal J}^i(\Psi)+{\cal J}_\Psi^i({\cal G}^a\Psi-\Delta^a)
= \intx ({\cal T}_{(\vf)}^{ai}\Psi-\Delta_{(\vf)}^{ai})
\nonumber\\
{\cal T}_{(\vfb)\,i}^a{\cal J}^j(\Psi)-{\cal J}^j_\Psi
({\cal T}_{(\vfb)\,i}^c\Psi-\Delta_{(\vfb)\,i}^c)=
\delta^j_i(\fud{\Psi}{\cb^a}+\partial_\mu\fud{\Psi}{K^a_\mu})
\nonumber\\
{\cal T}_{(\ob)\,i}^a {\cal F}^j\Psi
-{\cal F}^j({\cal T}_{(\ob)\,i}^a\Psi-\Delta_{(\ob)\,i}^a)=
-\delta_i^j (\fud{\Psi}{\cb^a}+\partial_\mu\fud{\Psi}{K^a_\mu})
\nonumber\\
{\cal R}^j_i{\cal S}(\Psi)+\BB_\Psi {\cal R}^j_i\Psi
={\cal U}^j_i\Psi
\nonumber\\
{\cal R}^j_i{\cal J}^k(\Psi)+{\cal J}^k_\Psi {\cal R}^j_i\Psi
=\delta^k_i {\cal F}^j\Psi
\nonumber\\
{\cal T}_{(\vf)}^{ai} {\cal R}^j_k\Psi -
{\cal R}^j_k({\cal T}_{(\vf)}^{ai}\Psi-\Delta_{(\vf)}^{ai})=
\delta^i_k({\cal T}_{(\omega)}^{aj}\Psi-\Delta_{(\omega)}^{aj})
\nonumber\\
{\cal T}_{(\ob)\,i}^a {\cal R}^j_k\Psi +
{\cal R}^j_k({\cal T}_{(\ob)\,i}^a\Psi-\Delta_{(\ob)\,i}^a)=
-\delta^j_i({\cal T}_{(\vfb)\,k}^{a}\Psi-\Delta_{(\vfb)\,k}^{a})
\ ,
\ea\eqn{algebra}
where $\Psi$ is a generic functional of even ghost charge and
${\cal J}^i_\Psi$ is the linearized operator corresponding to~\equ{315}.

The model we are considering turns out to be completely determined by
\begin{itemize}
\item the gauge condition~\equ{321} and the local
      equations~\equ{322}\ ;
\item the Slavnov identity~\equ{33}\ ;
\item the ghost equation~\equ{26}\ ;
\item the symmetry~\equ{311}\ ;
\item the quantum numbers listed in Table 1.
\end{itemize}

\section{Renormalization}

We prove the renormalizability of the model by first finding the most
general counterterm compatible with the algebraic structure described in
the previous section and then by
showing that the symmetries considered
hold to all orders of perturbation theory, {\it i.e.} that they are not
anomalous.

\subsection{Counterterm}

According to the Quantum Action Principle (QAP)~\cite{qap}, the
counterterm is the most general integrated local functional $\S_\Delta$
of dimension four
with vanishing ghost and $Q_f$ numbers satisfying the identities
\eq
\ba{rl}
\fud{\S_\Delta}{b^a} &= 0\\
\partial^\mu\fud{\S_\Delta}{K^{a\mu}} +& \fud{\S_\Delta}{\bar{c}^a}= 0\\
{\cal T}_{(\vf)}\Sigma_\Delta = {\cal T}_{(\vfb)}\Sigma_\Delta =&
{\cal T}_{(\omega)}\Sigma_\Delta = {\cal T}_{(\ob)}\Sigma_\Delta =0\ ,
\ea\eqn{41}
and
\eq
B_\S\S_\Delta=0\ ,
\eqn{42}
\eq
{\cal G}^a\S_\Delta=0\ ,
\eqn{43}
\eq
{\cal F}^i\S_\Delta=0\ .
\eqn{44}

The relations~\equ{41} imply~\cite{zw93} that $\S_\Delta$ is in fact only a
functional of
\eq
\S_\Delta\left[
A,c,\K,L,\M,\N,\U,\V \right]\ ,
\eqn{45}
where
\eq\ba{rl}
\K^a_\mu =& K^a_\mu +\partial_\mu \cb^a +
f^{abc}(U^{bi}_\mu + \partial_\mu\ob^{bi})\vf^c_i+f^{abc}V^b_{\mu
i}\ob^{ci}
\\
\M^{ai}_\mu =& M^{ai}_\mu + \partial_\mu\vfb^{ai}
\\
\N^a_{\mu i}=&N^a_{\mu i} + \partial_\mu\omega^a_i
\\
\U^{ai}_\mu=&U^{ai}_\mu + \partial_\mu\ob^{ai}
\\
\V^a_{\mu i}=&V^a_{\mu i} + \partial_\mu \vf^a_i
\ .
\ea\eqn{46}

The most general counterterm satisfying the Slavnov
condition~\equ{42} with ghost and $Q_f$ numbers zero therefore is
\eq
\S_\Delta=c_0\intx F^{a}_{\mu\nu}F^a_{\mu\nu} + B_\S \intx \LP
c_1\K^a_\mu A^a_\mu + c_2 \L^ac^a +c_3 \U^{ai}_\mu\V^a_{\mu i}\RP\ ,
\eqn{47}
where $c_0, c_1, c_2, c_3$ are arbitrary constants. Exploiting the ghost
condition~\equ{43}, we obtain
\eq
c_2=0\ ,
\eqn{48}
a reduction of the possible divergences peculiar for the Landau
gauge~\cite{landau}.
Finally the constraint implies that
\eq
c_1=-c_3\ .
\eqn{49}

We have thus algebraically shown that the model defined by the classical
action~\equ{24} and the symmetries~\equ{33},~\equ{26},~\equ{311} has two
divergences
\eq
\S_\Delta =c_0\intx F^{a}_{\mu\nu}F^a_{\mu\nu} + c_1B_\S \intx \LP
\K^a_\mu A^a_\mu - \U^{ai}_\mu\V^a_{\mu i}\RP\ ,
\eqn{410}
that can be absorbed through two independent multiplicative
renormalization constants, in complete agreement
with ordinary Yang Mills theory in the Landau gauge~\cite{book}.
The degrees of freedom introduced to constrain the functional integral to the
FMR~\cite{zw89,zw92} therefore do not lead to additional divergences.

\subsection{Anomalies}

Since the new fields introduced to constrain the functional integral to
the FMR all appear as BRS doublets, it is obvious that they
do not belong to the cohomology of the Slavnov operator~\cite{dixon},
which therefore  is cohomologically equivalent to
the ordinary one. Algebraically one only finds the usual
Adler-Bardeen anomaly, whose coefficient is known to vanish if all
fields transform according to  real representations of the gauge
group~\cite{adler}.
One still has to show that the other symmetries used to obtain the
result~\equ{410} are not anomalous as well.

Although the conventional procedure~\cite{brs} of individually implementing the
identities defining the model is quite straightforward in our case
thanks to the absence of gauge anomalies, we adopt here the technique of
collecting the symmetries into one nilpotent operator by introducing
global ghosts~\cite{d}. This method is particularly convenient when proving
that a whole algebra of operators is free of anomalies.

It is trivial to show that the
identities~\equ{321},~\equ{antighosteq},~\equ{322} hold for the quantum
vertex functional
\eq
\Gamma=\S+\Gamma^{(qu)}\ ,
\eqn{411}
where $\Gamma^{(qu)}$ is at least of order~$\hbar$. In section 4 we will
exploit that $\Gamma^{(qu)}$ is consequently a functional of the
 combinations~\equ{46} , $c$ and the connection $A$ only.


To collect the symmetries~\equ{33},~\equ{26},~\equ{39},~\equ{311}
and~\equ{314} into one operator, we
first consider the transformation on the fields generated by
\eq\ba{rl}
Q=&
s+\xi^a\GG^a+\eta^a\HH^a_{rig}+\lambda_i\FF_0^i
+\varrho_i\JJ_0^i
+\s^a_i\wT^{ai}_{(\o)}+\tau^a_i\wT^{ai}_{(\vf)}
\\&
-(\xi^a-\frac{1}{2}f^{abc}\eta^b\eta^c)
\pad{\ }{\eta^a}-\lambda_i\pad{\ }{\vr_i}
+f^{abc}\eta^b\xi^c\pad{\ }{\xi^a}
\\&
+(\xi^a\lambda_i+f^{abc}
(\eta^b\s^c_i-\xi^b\tau^c_i))\pad{\ }{\s^a_i}
-(\s^a_i+\xi^a\vr_i-f^{abc}\eta^b\tau^c_i)\pad{\ }{\tau^a_i}\ ,
\ea\eqn{412}
where
\eq
\FF^i_0=
\intx \LP c^a\fud{\ }{\omega^a_i}- \bar\omega^{ai}\fud{\ }{\bar{c}^a}\RP\ ,
\eqn{413}
\eq
\JJ^i_0=\intx\LP c^{a} \fud{\ }{{\vf^a_i}}
   -  (sc^a)\fud{\ }{\omega^a_i}
   +  \vfb^{ai}\fud{\ }{\bar{c}^a}\RP\ ,
\eqn{414}
\eq
\wT^{ai}_{(\o)}= \intx \LP\fud{\ }{\omega^a_i}
+ f^{abc}\ob^{bi}\fud{\ }{b^c}\RP\ ,
\eqn{wto}
\eq
\wT^{ai}_{(\vf)}= \intx\LP \fud{\ }{\vf^a_i}
+f^{abc}( \vfb^{bi}\fud{\ }{b^c}
+\ob^{bi}\fud{\ }{\cb^c})\RP\ .
\eqn{wtf}

In the definition~\equ{412} we introduced global ghost fields
$(\xi,\eta,\lambda,\vr,\s,\tau)$, whose quantum numbers are
summarized in Table 2.

\begin{center}
\begin{tabular}{|l|r|r|r|r|r|r|}\hline
&$\xi$&$\eta$&$\lambda$&$\vr$&$\s$&$\tau$
\\ \hline
dim&0&0&1&1&1&1
\\ \hline
$\Phi\Pi$&2&1&1&0&2&1
\\ \hline
$Q_f$&0&0&1&1&$-1$&$-1$
\\ \hline
\end{tabular}
\nobreak

\vspace{.2cm}{\footnotesize
{\bf Table 2.} Dimensions, Faddeev--Popov charges and $Q_f$ numbers of
the global ghosts.}
\end{center}

The operator $Q$ is nilpotent
\eq
Q^2=0\ .
\eqn{q20}
It does not describe a symmetry of the action $S$
\eq
QS=\intx\LP  (\vr_i M^{ai}_\mu-\lambda_iU^{ai}_\mu
+f^{abc}\tau^b_iU^{ci}_\mu) (D_\mu c)^a
-\tau^a_i(D_\mu M^i_\mu)^a
-\s^a_i(D_\mu U^i_\mu)^a\RP
\ ,
\eqn{415}
but the modified classical action
\eq
I=S+S_{ext}^{(Q)}\ ,
\eqn{416}
where
\eq\ba{rl}
S_{ext}^{(Q)}=\intx\LP&\!\! K^a_\mu (QA^a_\mu) + L^a (Qc^a) +
 X^{ai}(Q\omega^a_i)
 +(\vr_iM^{ai}_\mu-\lambda_iU^{ai}_\mu
+f^{abc}\tau^b_iU^{ci}_\mu)A^a_\mu\RP\ ,
\ea\eqn{417}
satisfies the generalized Slavnov identity
\eq
{\cal D}(I)=0\ ,
\eqn{418}
with
\eq\ba{rl}
{\cal D}(I)  =   \intx \LP &\!\!
       \fud{I}{K^a_\mu} \fud{I}{A^{a}_{\mu}}
   +   \fud{I}{L^{a}} \fud{I}{c^{a}}
   +    \fud{I}{X^{ai}} \fud{I}{\omega^{a}_i}
   +   (Qb^a)\fud{I}{b^a}
   +   (Q\cb^a)\fud{I}{{\bar c}^{a}}
\\&
   +   (Q\vf^a_i) \fud{I}{\vf^{a}_i}
   +    (Q\vfb^{ai}) \fud{I}{\bar\vf^{ai}}
   +    (Q\ob^{ai}) \fud{I}{\bar\omega^{ai}}
\\&
   +   (QM^{ai}_\mu) \fud{I}{M^{ai}_\mu}
   +   (QN^a_{\mu i}) \fud{I}{N^a_{\mu i}}
   +   (QU^{ai}_\mu) \fud{I}{U^{ai}_\mu}
   +   (QV^a_{\mu i}) \fud{I}{V^a_{\mu i}}\RP
\\&
-(\xi^a-\frac{1}{2}f^{abc}\eta^b\eta^c)
\pad{I}{\eta^a}-\lambda_i\pad{I}{\vr_i}+
f^{abc}\eta^b\xi^c\pad{I}{\xi^a}
\\&
+(\xi^a\lambda_i+f^{abc}
(\eta^b\s^c_i-\xi^b\tau^c_i))\pad{I}{\s^a_i}
-(\s^a_i+\xi^a\vr_i-f^{abc}\eta^b\tau^c_i)\pad{I}{\tau^a_i}\ .
\ea
\eqn{419}

The corresponding linearized operator
\eq\ba{rl}
{\cal D}_I  =   \intx \LP &\!\!
       \fud{I}{K^a_\mu} \fud{\ }{A^{a}_{\mu}}
+      \fud{I}{A^a_\mu} \fud{\ }{K^{a}_{\mu}}
   +   \fud{I}{L^{a}} \fud{\ }{c^{a}}
   +   \fud{I}{c^{a}} \fud{\ }{L^{a}}
   +    \fud{I}{X^{ai}} \fud{\ }{\omega^{a}_i}
   +    \fud{I}{\omega^a_i} \fud{\ }{X^{ai}}
\\&
   +   (Qb^a)\fud{\ }{b^a}
   +   (Q\cb^a)\fud{\ }{\cb^a}
   +   (Q\vf^a_i) \fud{\ }{\vf^{a}_i}
   +    (Q\vfb^{ai}) \fud{\ }{\bar\vf^{ai}}
   +    (Q\ob^{ai}) \fud{\ }{\bar\omega^{ai}}
\\&
   +   (QM^{ai}_\mu) \fud{\ }{M^{ai}_\mu}
   +   (QN^a_{\mu i}) \fud{\ }{N^a_{\mu i}}
   +   (QU^{ai}_\mu) \fud{\ }{U^{ai}_\mu}
   +   (QV^a_{\mu i}) \fud{\ }{V^a_{\mu i}}\RP
\\&
-(\xi^a-\frac{1}{2}f^{abc}\eta^b\eta^c)
\pad{\ }{\eta^a}-\lambda_i\pad{\ }{\vr_i}+
f^{abc}\eta^b\xi^c\pad{\ }{\xi^a}
\\&
+(\xi^a\lambda_i+f^{abc}
(\eta^b\s^c_i-\xi^b\tau^c_i))\pad{\ }{\s^a_i}
-(\s^a_i+\xi^a\vr_i-f^{abc}\eta^b\tau^c_i)\pad{\ }{\tau^a_i}\ ,
\ea
\eqn{421}
is nilpotent
\eq
\DD_I\DD_I=0\ .
\eqn{dnil}

The introduction of the global ghosts leads to the following identities
for the action~$I$
\eq\ba{rl}
\pad{I}{\xi^a}=\Delta^a_{(\xi)}\ \ ;\ \
\pad{I}{\eta^a}=\Delta^a_{(\eta)}\ \ ;\ \
\pad{I}{\lambda_i}=\Delta^i_{(\lambda)}\ \ ;\ \
\pad{I}{\s^a_i}=&\Delta^{ai}_{(\s)}
\ \ ;\ \
\pad{I}{\tau^a_i}=\Delta^{ai}_{(\tau)}
\\
\TT^i_{(\vr)}I\equiv\pad{I}{\vr_i}+\intx
X^{ai}\fud{I}{L^a}=&\Delta^i_{(\vr)}\ ,
\ea\eqn{422}
where
\eq\ba{rl}
\Delta^a_{(\xi)}=&\intx \LP L^a-f^{abc}X^{bi}\vf^c_i\RP\\
\Delta^a_{(\eta)}=&\intx f^{abc}\LP
K^b_\mu A^c_\mu - L^bc^c - X^{bi}\omega^c_i\RP\\
\Delta^i_{(\lambda)}=&\intx\LP X^{ai}c^a
- U^{ai}_\mu A^a_\mu\RP\\
\Delta^i_{(\vr)}=&\intx M^{ai}_\mu A^a_\mu\\
\Delta^{ai}_{(\s)}=&\intx X^{ai}\\
\Delta^{ai}_{(\tau)}=&\intx f^{abc}U^{bi}_\mu A^c_\mu \ ,
\ea\eqn{423}
are linear breakings.\\
The nonlinear algebra, valid for any even ghost charged functional~$\Psi$
\eq\ba{rcl}
\pad{\ }{\xi^a}\DD(\Psi)&-&\DD_\Psi(\pad{\Psi}{\xi^a}
- \Delta^a_{(\xi)}) =
(G^a\Psi-\Delta^a)
-(\pad{\Psi}{\eta^a}-\Delta^a_{(\eta)})
-f^{abc}\eta^b(\pad{\Psi}{\xi^c}-\Delta^c_{(\xi)})
\\&+&
\lambda_i(\pad{\Psi}{\s^a_i}-\D^{ai}_{(\s)})
-\vr_i(\pad{\Psi}{\tau^a_i}-\D^{ai}_{(\tau)})
-f^{abc}\tau^b_i(\pad{\Psi}{\s^c_i}-\D^{ci}_{(\s)})
\\&-&
\intx\LP f^{abc}(\vr_iX^{bi}c^c + \vr_i
U^{bi}_\mu A^c_\mu + \eta^bL^c -\vf^b_if^{cde}X^{di}\eta^e)
-\lambda_iX^{ai}\RP
\ea\eqn{al1}
\eq\ba{rcl}
\pad{\ }{\eta^a}\DD(\Psi)&+&\DD_\Psi(\pad{\Psi}{\eta^a}
- \Delta^a_{(\eta)}) =
\HH^a_{rig}\Psi
\\&+&
f^{abc}\eta^b(\pad{\Psi}{\eta^c}-\Delta^c_{(\eta)})
+f^{abc}\xi^b(\pad{\Psi}{\xi^c}-\Delta^c_{(\xi)})
+f^{abc}\s^b_i(\pad{\Psi}{\s^c_i}-\Delta^{ci}_{(\s)})
\\&+&
f^{abc}\tau^b_i(\pad{\Psi}{\tau^c_i}-\Delta^{ci}_{(\tau)})
+f^{abc}(\eta^b\Delta^c_{(\eta)}+\xi^b\Delta^c_{(\xi)}
+\s^b_i\D^{ci}_{(\s)}+\tau^b_i\D^{ci}_{(\tau)})
\ea\eqn{al2}
\eq\ba{rcl}
\pad{\ }{\lambda_i}\DD(\Psi)&+&
\DD_\Psi(\pad{\Psi}{\lambda_i}- \Delta^i_{(\lambda)}) =
\FF^i\Psi
-(\TT^i_{(\vr)}\Psi-\Delta^i_{(\vr)})
+\xi^a(\pad{\Psi}{\s^a_i}-\D^{ai}_{(\s)})
\\&+&
\intx(\xi^aX^{ai}+f^{abc}\eta^aU^{bi}_\mu A^c_\mu)
\ea\eqn{al3}
\eq
\TT_{(\vr)}^i\DD(\Psi)-\DD_\Psi
(\TT_{(\vr)}^i\Psi-\Delta^i_{(\vr)})=\JJ^i(\Psi)
-\xi^a(\pad{\Psi}{\tau^a_i}-\D^{ai}_{(\tau)})
+\intx f^{abc}\eta^aM^{bi}_\mu A^c_\mu
\eqn{al4}
\eq\ba{rcl}
\pad{\ }{\s^a_i}\DD(\Psi)&-&\DD_\Psi(\pad{\Psi}{\s^a_i}-\D^{ai}_{(\s)})=
\intx (T^{ai}_{(\o)}\Psi-\D^{ai}_{(\o)})
-(\pad{\Psi}{\tau^a_i}-\D^{ai}_{(\tau)})
\\&-&
f^{abc}\eta^b(\pad{\Psi}{\s^c_i}-\D^{ci}_{(\s)})
-f^{abc}\eta^b\D^{ci}_{(\s)}
\ea\eqn{al5}
\eq\ba{rcl}
\pad{\ }{\tau^a_i}\DD(\Psi)&+&\DD_\Psi
(\pad{\Psi}{\tau^a_i}-\D^{ai}_{(\tau)})=
\intx (T^{ai}_{(\vf)}\Psi-\D^{ai}_{(\vf)})
+f^{abc}\eta^b(\pad{\Psi}{\tau^c_i}-\D^{ci}_{(\tau)})
\\&+&
f^{abc}\xi^b(\pad{\Psi}{\s^c_i}-\D^{ci}_{(\s)})
+\intx f^{abc}(\xi^bX^{ci}+U^{bi}_\mu f^{cde}\eta^dA^e_\mu)\ ,
\ea\eqn{al6}

implies for a functional $\Psi=\Gamma^{(Q)}$ satisfying
\eq
\pad{\ }{\xi^a}\Gamma^{(Q)}=\Delta^a_{(\xi)}\ \ ;\ \
\pad{\ }{\eta^a}\Gamma^{(Q)}=\Delta^a_{(\eta)}\ \ ;\ \
\pad{\ }{\lambda_i}\Gamma^{(Q)}=\Delta^i_{(\lambda)}
\eqn{425-1}
\eq
\TT^i_{(\vr)}\Gamma^{(Q)}=\Delta^i_{(\vr)}\ \ ;\ \
\pad{\ }{\s^a_i}\G^{(Q)}=\D^{ai}_{(\s)}\ \ ;\ \
\pad{\ }{\tau^a_i}\G^{(Q)}=\D^{ai}_{(\tau)}
\eqn{425-2}
\eq
\DD(\Gamma^{(Q)})=0\ ,
\eqn{426}

that the following identities hold
\eq\ba{rl}
G^a\Gamma^{(Q)}=&\Delta^a
+\intx ( f^{abc}(\vr_iX^{bi}c^c + \vr_i
U^{bi}_\mu A^c_\mu + \eta^bL^c -\vf^b_if^{cde}X^{di}\eta^e)
-\lambda_iX^{ai})
\\
\HH^a_{rig}\Gamma^{(Q)}=&
-f^{abc}(\eta^b\Delta^c_{(\eta)}+\xi^b\Delta^c_{(\xi)}
+ \s^b_i\D^{ci}_{(\s)}+\tau^b_i\D^{ci}_{(\tau)})
\\
\FF^i\Gamma^{(Q)}=&
-\intx(\xi^aX^{ai}+f^{abc}\eta^aU^{bi}_\mu A^c_\mu)
\\
\JJ^i(\Gamma^{(Q)})=&
-\intx f^{abc}\eta^a
M^{bi}_\mu A^c_\mu
\\
\intx T^{ai}_{(\o)}\Gamma^{(Q)}=&\intx (\D^{ai}_{(\o)}
+ f^{abc}\eta^b\D^{ci}_{(\s)})
\\
\intx T^{ai}_{(\vf)}\Gamma^{(Q)}
=&\intx (\D^{ai}_{(\vf)}
- f^{abc}(\xi^bX^{ci}+U^{bi}_\mu f^{cde}\eta^dA^e_\mu))
\ .
\ea\eqn{427}

 From equations~\equ{418},~\equ{427} one sees that at vanishing global
ghosts and source $X$ the quantum vertex functional
$\left.\Gamma\equiv\Gamma^{(Q)}\right|_{\xi=\eta=\lambda=\rho=X=0}$
satisfies
\eq\ba{rl}
\SS(\Gamma)=&0\\
\GG^a\Gamma=&\Delta^a\\
\HH^a_{rig}\Gamma=&0\\
\FF^i\Gamma=&0\\
\JJ^i(\Gamma)=&0\\
\intx T^{ai}_{(\o)}\G=&\intx\D^{ai}_{(\o)}\\
\intx T^{ai}_{(\vf)}\G=&\intx\D^{ai}_{(\vf)}\ ,
\ea
\eqn{428}
which is the desired result.

It is quite straightforward to show that the relations~\equ{425-1},~\equ{425-2}
hold and it is apparent from the nonlinear algebra~\equ{al1}--\equ{al6}
that proving the absence of anomalies  has been reduced to
showing that the generalized Slavnov identity~\equ{418} is not anomalous.
We can now apply the mathematical tools developed
for nilpotent operators~\cite{dixon}.

 From the QAP~\cite{qap} we know that the generalized Slavnov identity could be
broken at quantum level
\eq
\DD(\Gamma^{(Q)})=\AA\cdot\Gamma^{(Q)}\ ,
\eqn{429}
only by a quantum insertion $\AA\cdot\Gamma^{(Q)}$, which to
lowest order in~$\hbar$ is an integrated local functional of dimension
four, ghost charge +1 and $Q_f$ counting number zero
\eq
\AA\cdot\Gamma^{(Q)}=A+O(\hbar A)\ .
\eqn{430}
This lowest order breaking $A$ must
satisfy the Wess--Zumino consistency condition~\cite{zumino,bannals}
\eq
\DD_I A=0 \ .
\eqn{431}
Since $\DD_I$ is a nilpotent operator, eq.~\equ{431} is a cohomology
problem that we solve by decomposing $\DD_I$ with the filtration
operator~\cite{dixon}
\eq
\NN=\xi^a\pad{\ }{\xi^a}+\eta^a\pad{\ }{\eta^a}+
\lambda_i\pad{\ }{\lambda_i}+
\vr_i\pad{\ }{\vr_i}+\sigma_i^a\pad{\ }{\sigma_i^a}
+\tau_i^a\pad{\ }{\tau_i^a}
\eqn{432}
into
\eq
\DD_I=\DD^{(0)}+\DD^{(R)}\ ,
\eqn{433}
where
\eq
D^{(0)}=\BB_\S-\xi^a\pad{\ }{\eta^a}-\lambda_i\pad{\ }{\vr_i}
-\sigma_i^a\pad{\ }{\tau_i^a}
\eqn{434}
Because of~\equ{36}, the operator $D^{(0)}$ is nilpotent, and the
result of~\cite{dixon} ensures that the
cohomology of $\DD_I$ is isomorphic to a subspace of that of $D^{(0)}$,
which does not depend on the global ghosts
$(\eta,\xi;\vr,\lambda;\tau,\sigma)$ nor on the fields $(\vf,\omega;\ob,
\vfb;U,M;V,N)$ since they appear in~\equ{434} as BRS
doublets~\cite{dixon}. We are therefore
left to study the cohomology problem
\eq
B_\S X=0\ ,
\eqn{435}
where $B_\S $ is the linearized Slavnov operator of ordinary Yang--Mills
theory. As discussed previously, the solution of~\equ{435} is a trivial
cocycle since there is no Adler-Bardeen anomaly in this
model~\cite{adler}, and consequently the cohomology of~$\DD_I$ is empty.

We have thus proved that the solution of the Wess--Zumino consistency
condition~\equ{431} is
\eq
A=\DD_I\widehat A\ ,
\eqn{452}
{\it i.e.} that the generalized Slavnov identity~\equ{418} is not
anomalous, and
that the symmetries~\equ{428} we considered are therefore valid to all
orders of perturbation theory.

Along the same lines, it is straightforward to also prove that the
symmetries~\equ{316}
and~\equ{319} are anomaly--free by starting from the
transformations generated by
the nilpotent operator
\eq
Q^\prime = s +\lambda^j_i {\RR}^i_j +\vr^j_i {\UU}_j^i
-(\lambda^i_j + \vr^i_k \vr^k_j)\pad{\ }{\vr^i_j}
-(\lambda^k_j \vr_k^i - \lambda^i_k \vr^k_j)\pad{\ }{\lambda^i_j}\ ,
\eqn{}
where $(\lambda^i_j, \vr^i_j)$ are again global ghosts.

All the symmetries that form the algebra~(2.33) are thus valid at
the quantum level, and the unitarity of the model is ensured~\cite{bannals}.

\section{The model at physical sources}

The analysis of the Gribov ambiguity made in Ref.~\cite{zw89,zw92}, has
demonstrated that the functional integration is effectively
constrained to the FMR in a quantum theory defined by the classical
action~\equ{24} for nonvanishing physical sources
\eq
M^{a}_{\mu\nu b}\left.\right|_{{\rm ph}}=
- V^{a}_{\mu\nu b}\left.\right|_{{\rm
ph}}=\gamma\delta_{\mu\nu}\delta^a_b\ ,
\eqn{51}
where the mass parameter is determined selfconsistently by the horizon
condition~\cite{zw93}
\eq
\left.\pad{\Gamma}{\gamma}\right|_{{\rm ph}}=0\ ,
\eqn{52}
when the quantum fields
$\Phi\in\{A,c,\cb,b,\vf,\vfb,\omega,\ob\}$
assume their  vacuum values
\eq
\left.\fud{\Gamma}{\Phi}\right|_{\Phi=\Phi|_{{\rm ph}}}=0 \ .
\eqn{53}

As shown in~\cite{zw92}, equation~\equ{52} can only be fulfilled at a
nonvanishing value for~$\gamma$.

At their physical values~\equ{51} the sources do not appear as $BRS$
doublets and the classical action is no longer $BRS$-symmetric
\eq
s\LP\left. S\right|_{{\rm ph}}\RP = \gamma s (D_\mu\vf_{\mu a})^a\ .
\eqn{54}
To show that this term can be interpreted as arising from spontaneous
symmetry breakdown, consider first the {\it symmetric} quantum vertex
functional at vanishing sources
\eq
\Gamma^{sym}=\Gamma|_{M=N=U=V=0}\ .
\eqn{s0}
In the previous sections we have shown that it is a finite functional of
the renormalized fields and coupling constant. The replacements
\eq
\vfb^{a}_{\mu b}\Longrightarrow \vfb^{\prime a}_{\mu b}
+\gamma_Mx_\mu\delta^a_b
\eqn{55}
\eq
\vf^{a}_{\mu b}\Longrightarrow \vf^{\prime a}_{\mu b}
-\gamma_Vx_\mu\delta^a_b
\eqn{56}
\eq
\cb^a\Longrightarrow \cb^{\prime a}-\gamma_Vf^{abc}x_\mu \ob^{c}_{\mu b}
\eqn{58}
\eq
b^a\Longrightarrow b^{\prime a}-\gamma_V f^{abc}\vfb^{\prime c}_{\mu b}x_\mu
\eqn{59}
lead to a quantum vertex functional of the shifted quantum fields, which
is the one for nonvanishing external sources\footnote[1]{D. Zwanziger observed
that the shifts~\equ{55}--\equ{59}  also eliminate the
$BRS$--breaking terms in the lattice regularized version of this
model~\cite{zwpreprint}.}

\eq
\Gamma^{sym}[\vf,\vfb,\omega,\ob,c,\cb,b,A]=
\Gamma[\vf^\prime,\vfb^\prime,\omega,\ob,c,\cb^\prime,b^\prime,A;
M^\prime,V^\prime]\ ,
\eqn{relgam}
with
\eq
M^{\prime a}_{\mu\nu b} = \gamma_M\delta_{\mu\nu}\delta^a_b\ \ \, \ \ \ \
V^{\prime a}_{\mu\nu b} =
-\gamma_V\delta_{\mu\nu}\delta^a_b\  .
\eqn{shiftscource}
Relation~\equ{relgam} can easily be verified from the form of the
classical action~\equ{24} and the fact that the radiative correction
$\Gamma^{(qu)}$  in~\equ{411}
only depends on the combinations~\equ{46}.
We obtain the quantum vertex functional at the physical value of the
sources~\equ{51} if we set
\eq
\gamma_V=\gamma_M=\gamma\ .
\eqn{510}

As in theories with spontaneously broken symmetries, we find that the
quantum theory is also renormalizable in the asymmetric case, because
it is equivalent to introducing non-vanishing external sources.

It is remarkable that the explicit coordinate dependence of the
shifts~\equ{55}--\equ{59} is not reflected in $\Gamma$. This can be
traced to the invariance of the symmetric quantum vertex functional
under the global $U(f)$ group in addition to its $O(4)$ and $SU(N)$
symmetry under euclidean coordinate-- and rigid gauge-- transformations.
Each of these symmetries is individually broken spontaneously by the
shifts~\equ{55}--\equ{59} but a diagonal $SU(N)\times O(4)$ subgroup
remains intact, which assures coordinate and global
colour invariance also in the broken phase.

The
analogy with spontaneous symmetry breakdown can be further pursued,
because the shifts~\equ{55}--\equ{59} also change the vacuum values of
the quantum fields. Perturbation theory with physical
values~\equ{41} of the sources corresponds to an expansion around
nontrivial vacuum values in the symmetric theory. Furthermore
the difference in the classical vacuum energy density between the
trivial and nontrivial vacuum values~\equ{55}--\equ{59}
\eq
\Delta\epsilon_{vac}=
\frac{1}{V}\LP
\left. \S^{sym}\right|_{\vfb^\prime=\vf^\prime=\omega^\prime=
\cb^\prime=b^\prime=\ob=A=c=0} - \left. \S^{sym}\right|_{\vfb=\vf=\omega=
\cb=b=\ob=A=c=0}\RP\! = -4(N^2-1)\gamma_V\gamma_M ,
\eqn{last}
with $\S^{sym}=\S|_{M=N=U=V=0}$, implies that the
 broken phase is energetically
preferred. We can interpret the horizon equation~\equ{52} as a
minimizing condition for the vacuum energy density in the
presence of quantum fluctuations. These are in fact necessary to
satisfy~\equ{52}, because~\equ{last} only depends linearly on
$\gamma_M\gamma_V=\gamma^2$.

Although surprising and perhaps even disturbing, a careful analysis of
covariant gauge fixing on the lattice also indicated that the BRS-symmetry
could be spontaneously broken, in order to avoid that the summation over Gribov
copies conspires to yield vanishing expectation values for gauge invariant
observables~\cite{Neuberger}.

\section{Conclusions}

The model defined by the classical action~\equ{24} was proposed~\cite{zw93}
to effectively
restrict the functional integration of Yang-Mills theories to the FMR~\equ{114}
by means of additional fields and external
sources which satisfy a selfconsistency or
``horizon'' condition~\equ{51}--\equ{53} at the physical point.

This restriction to the FMR is a refinement of the usual Landau gauge, a kind
of $minimal$ one
without Gribov copies. We indeed recovered also for this model the
property~\cite{landau} of Landau gauges that the
integrated ghost equation of motion yields a Ward Identity (equ.~\equ{26}).
The rich symmetry structure~(2.33)  allowed us to prove algebraically
that only two independent divergences appear in a
perturbative analysis, which means that the model has the same
renormalization properties as ordinary Yang Mills theory in Landau
gauge~\cite{book} in spite of the additional fields and sources. The
renormalization proof was completed by showing that the  symmetries of the
model do hold to all orders of perturbation theory, i.e. are
not anomalous. Unitarity of the physical S-matrix is then a consequence
of the validity to all orders of perturbation theory of the
Slavnov identity.

We believe that the algebraic structure of the
enlarged theory effectively eliminates all the additional degrees of
freedom introduced. This would be in the spirit which led to
the construction of the model, namely constraining the gauge field
configurations to the FMR, without altering the physical content of the
original Yang Mills theory~\cite{zw89,zw92}. This conjecture is supported by
the observation that the $BRS$ breaking term~\equ{54} at the physical point
can be understood as resulting from the nonperturbative
shifts~\equ{55}--\equ{59}  in the $BRS$--symmetric case and that the horizon
condition minimizes the vacuum energy density.

\vspace{2cm}
\begin{center}
\bf Acknowledgements
\end{center}
We would like to thank D. Zwanziger for stimulating discussions and
helpful comments.

\end{document}